\documentclass[12pt]{article}
\usepackage[T1]{fontenc}
\usepackage[latin1]{inputenc}
\usepackage{epsfig,graphics,cite}

\providecommand{\LyX}{L\kern-.1667em\lower.25em\hbox{Y}\kern-.125emX\@}

\begin{document}

\date{\today}

\begin{center}
\textbf{\Large Dual coherent particle emission as generalized two-component
Cherenkov-like effect }
\end{center}

\smallskip

\begin{center}
{\large D. B. Ion}\\
{\small TH Division, CERN, CH-1211 Geneva 23, Switzerland,\\
and  NIPNE-HH,  Bucharest, P.O. Box MG-6, Romania }
\end{center}

\begin{center}
{\large E. K. Sarkisyan}\\
{\small EP Division, CERN, CH-1211 Geneva 23, Switzerland, and \\
Department of Physics, The University of Manchester, Manchester M13 9PL, UK}
\end{center}

\medskip 
\medskip

\begin{abstract}
In this paper we introduce a new kind of coherent particle production
mechanism called \emph{dual coherent particle emission (DCPE)} as
generalized two-component Cherenkov-like effect, which takes place when the
phase velocity of emitted particle $v_{Mph}$ and the particle source phase
velocity $v_{B_1ph}$ satisfy a specific DCPE condition: $v_{Mph}\leq 
v_{B_1ph}^{-1}$. The general signatures of the DCPE in dielectric, nuclear and
hadronic media are established and some experimental evidences are presented.
\newline
\end{abstract}

Cherenkov radiation (CR) was first observed in the early 1900's by the
experiments developed by Mary and Pierre Curie when studying radioactivity
emission. However, the nature of such radiation was unknown until the
experimental works (1934-1937) of P.A. Cherenkov and the theoretical
interpretation by I.E. Tamm and I.M. Frank (1937) \cite{0}. Then, it was
discovered that this phenomenon is produced by charged particles movings
with superluminal speeds in medium. Now, the CR is one of the cornerstones
of classical and quantum electrodynamics and it is the subject of many
studies related to the extension to other coherent particle emission via
Cherenkov-like effects \cite{2,4,2a,14,3,14a}.  The generalized
Cherenkov-like effects based on four fundamental interactions  has been
investigated and classified recently in \cite{3}. In particular, this
classification includes the nuclear (mesonic, $\gamma$,  weak
boson)-Cherenkov-like radiations as well as the high energy component  of
the coherent particle emission via (baryonic, leptonic, fermionic)
Cherenkov-like effects.  It is important to underline that the CR is
extensively used in experiments for counting and identifying relativistic
particles.  We note that, by the experimental observation of the
subthreshold \cite{1} and anomalous \cite{1a} CR,  it was clarified that
some of fundamental  aspects of the CR are still open. Therefore, more
theoretical and experimental investigations on CR as well  on the
generalized Cherenkov-like effects are needed.

\begin{figure}[t]
\center
\rotatebox{-90}{\ {\includegraphics[bb=125 20 487
772,width=6.4cm]{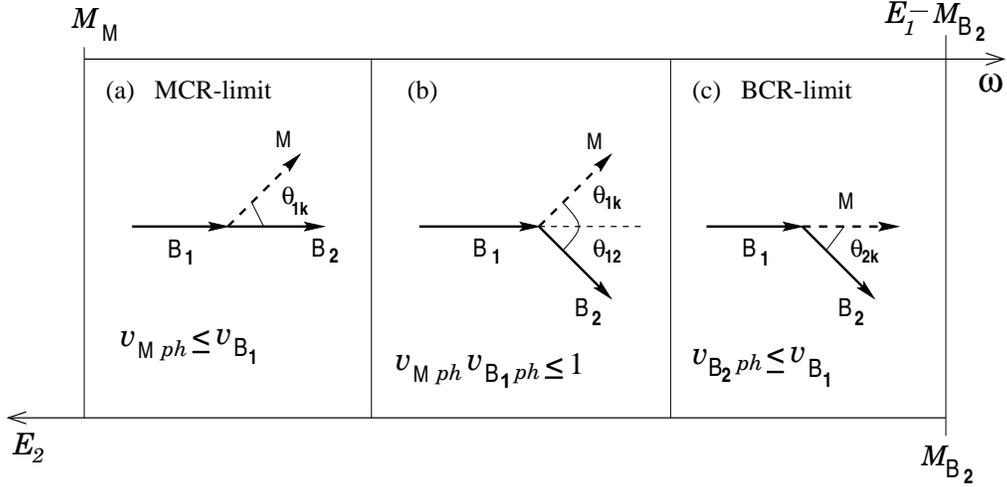}} } {}
\caption{ The schematic description of DCPE phenomenon as the generalized
two-component Cherenkov-like effect: (a) mesonic Cherenkov-like radiation
(MCR) limit \protect\cite{3}: $v_{Mph}\leq v_{B_1}$; (b) two-body decay
process B$_{1}$ $\rightarrow \mathrm{MB}_2$; (c) baryonic Cherenkov-like
radiation (BCR) limit \protect\cite{3}: $v_{B_{2}ph}\leq v_{B_1} $. }
\label{fig:1}
\end{figure}

In this paper we introduce a new kind of \textit{coherent} particle
production mechanism  in the medium called \emph{dual coherent particle
emission}  (DCPE) which includes in a more general and exact form not only
the usual  CRs but also all kind of the generalized Cherenkov-like  effects
and unifies them as \textit{two-component} Cherenkov-like effects. The
DCPE-effect is expected to take place when the \textit{phase velocities of
the emitted particle, $v_{Mph} $ (or $v_{B_2ph}$), and that of particle
source $v_{B_1ph} $ satisfy the dual coherence condition: 
$v_{B_1ph}v_{Mph}\leq 1 $} ($v_{B_1ph}v_{B_2ph}\leq 1 $). The 
general
signatures of DCPE effects in dielectric, nuclear and hadronic  media are
discussed.  For illustration some recent theoretical predictions and
experimental results  on the coherent meson production via mesonic
Cherenkov-like radiation in  nuclear \cite{5,6} and in hadronic \cite
{9,16,11} media are included. 
\bigskip

\noindent \emph{\textsc{Dual coherent particle emission}}. Let us start with
a general B$_{1}\rightarrow \mathrm{MB} $$_{2} $ decay, see Fig. \ref{fig:1}
(a). Here  a particle M [with energy $\omega , $ momentum $k=$ Re$
n_{M}(\omega ) $$\sqrt{\omega ^2-M^2_{\mathrm{M}}}$, rest mass $M_{\mathrm{M}
} $ and refractive index $n_{M}(\omega )] $ is emitted in a (nuclear,
hadronic, dielectric, etc.) medium by incident particle B$_{1} $ [with
energy $E_1 $, momentum $p _{1}=\mathrm{Re} n(E_{1})\sqrt{E^{2}_{1}-M_{1}^{2}
} $, rest mass M$_{1} $, refractive index $n_{1}(E_{1}) ${]} that itself
goes over into a final particle B$_{2} $ [with energy $E_{2} $, momentum $p
_2=\mathrm{Re}n(E_{2})\sqrt{E^{2}_{2}-M_{2}^{2}} $, rest mass $M_{2} $, 
refractive index $n_{2}(E_{2}) ${]}.

Here we prove that in order to obtain a genuine spontaneous particle
emission in a given medium \textit{the two general conditions} are necessary
to be fulfilled:

(i) \hangindent=1.cm \hangafter=1 {$\!\!$ The incident particle-source must
be coupled to a specific radiation field (RF) (see Fig. \ref{fig:1}) and
that the particles propagation properties in medium must be modified. }
\smallskip

\noindent (ii.1) \hangindent=.95cm \hangafter=1 {$\!$ The particle source
must be moving in the medium with a dual phase velocity $v^{-1}_{B_{1}ph} $
higher than the phase velocity $v_{Mph} $ of the RF-quanta.} \smallskip

\noindent (ii.2) \hangindent=.95cm \hangafter=1 {$\!$ The particle source
must be moving in medium with a dual phase velocity $v^{-1}_{B_{1}ph} \geq
v_{B_{2}ph} $. } \smallskip

\emph{Proof:} The propagation properties of particles in a medium are
changed in agreement with their elastic scattering with the constituents of
that medium. So, the phase velocity $v_{Xph}(E_X) $ of any particle X (with
the total energy $E_X$ and rest mass $M_{X} $) in medium is modified: 
\begin{equation}  \label{Eqo}
v_{Xph}(E_{X})=\frac{E_{X}}{p_{X}}=\frac{1}{\mathrm{Re}n_{X}(E_{X})}\cdot 
\frac{E_{X}}{\sqrt{E_{X}^{2}-M^{2}_{X}}}.
\end{equation}
The \emph{refractive index} $n_{X}(E_X) $ in a medium composed from the
constituents ``c'' can be calculated in standard way by using the Foldy-Lax
formula \cite{12} (we work in the units system ${\hbar}=c=1 $) 
\begin{equation}  
\label{Eq0}
n^{2}_{X}(E_{X})=1+\frac{4\pi \rho }{E_{X}^{2}-M^{2}_{X}}\cdot C(E_{X}) 
\overline{f}_{Xc\rightarrow Xc}(E_{X}),
\end{equation}
where $\rho $ is the density of the constituents, $C(E_{X})$ is a 
\emph{ coherence factor} [$C(E_{X})=1$ when the medium constituents are 
randomly
distributed], $\overline{f}_{Xc\rightarrow Xc}(E_X) $ is the averaged
forward Xc-scattering amplitude.

Now, by using the \emph{energy-momentum conservation law} for the decay $
\mathrm{B}_1 \rightarrow \mathrm{MB}_{2} $ in medium, $
E_{1}=E_{2}+\omega$, $\overrightarrow{p}_{1}=\overrightarrow{p}_{2}+
\overrightarrow{k}$, we obtain (see angles definition in Fig. \ref{fig:1}): 
\begin{equation}  \label{Eq10a}
\cos\theta _{1k}=v_{Mph}v_{B_{1}ph}+\frac{1}{2p_{1}k}(
-D_{B_{1}}+D_{B_{2}}-D_{M}),
\end{equation}
\vspace*{-.4cm} 
\begin{equation}  \label{Eq10c}
\cos\theta _{12}=v_{B_{1}ph}v_{B_{2}ph}+\frac{1}{2p_{1}p_2}(-D_{B_{1}}
-D_{B_{2}}+D_{M}),
\end{equation}
\vspace*{-.4cm} 
\begin{equation}  \label{Eq10b}
\cos\theta _{2k}= v_{Mph}v_{B_{2}ph}+\frac{1}{2p_{2}k}(
D_{B_{1}}-D_{B_{2}}-D_{M}).
\end{equation}
Here $D_{X} $,$\; X\equiv B_{1},B_{2},M, $ are departures off  mass shell, 
\[
D_{X}\equiv E^{2}_{X}-p^{2}_{X}=M^{2}_{X}+[1-(\mathrm{Re}
n_{X}(E_{X}))^{2}](E_{X}^{2}-M^{2}_{X}). 
\]

We note that the second terms in the right side of Eqs. (\ref{Eq10a})--(\ref
{Eq10b}) can be considered as quantum corrections to the first ``classical''
terms \cite{3}.

A rigorous proof of statement (ii) is obtained from the fact that the
respective emission angles must be the physical ones. The \emph{coherence
quantum conditions,} $\cos \theta _{ij} $ $\leq 1, $ $i,j=1,2,k,$ Eqs. (\ref
{Eq10a})--(\ref{Eq10b}), at high incident particle energies transform into 
\emph{classical coherence conditions}, e.g., 
\begin{equation}  \label{Eq1d}
\cos\theta _{1k}=v_{Mph}v_{B_{1}ph}\leq 1,
\end{equation}
which is equivalent to 
\begin{equation}  
\label{Eq1e}
v_{Mph}\leq v^{-1}_{B_{1}ph}\; \: \: \mathrm{or}\; \: \: v_{B_{1}ph}\leq
v^{-1}_{Mph}.
\end{equation}

It is worth to note that from the dual coherence conditions (\ref{Eq1e}) we
obtain the following \textit{limits of the DCPE condition}: (iii) \smallskip
In case when $v^{-1}_{B_{1}ph} =v_{B_1} $, from the condition (ii) the two
important generalized Cherenkov-like limits follow: (iii.1) the MCR limit $
v_{Mph} \leq v_{B_1} $, and (iii.2) the BCR limit $v_{B_{2}ph}\leq v_{B_1} $
(see Fig. 1)

The proof of the statement (iii) is obtained immediately if one observes
that \emph {when the particle $B_{1} $ is on the mass shell in medium 
(Re$n_1=1$), then $v^{-1}_{B_{1}ph} =v_{B_1} $,} and the \emph{
dual coherence conditions} (ii.1) and (ii.2) are the \emph{Cherenkov-like
coherence conditions} (iii.1) and (iii.2), respectively.

Now, we can obtain a classification of these DCPE effects not only on the
basis of four fundamental (strong, electromagnetic, weak, and gravitational)
interactions but also using the above \emph{``M-B duality''} as well as the 
\emph{crossing symmetry.}

The main signatures of the DCPE as the generalized two-component
Cherenkov-like effects are as follows: \smallskip

\begin{itemize}
\item  The differential cross sections posses the bumps in the energy bands
where the DCPE conditions are fulfilled.

\item  The DCPE-effects are threshold mechanisms. In the classical limit the
DCPE-threshold velocity is related with Cherenkov-like threshold velocity
via the relation $v_{B_{1}}^{thr}(DCPE)=v_{B_{1}}^{thr}(XCR)/\mathrm{Re}
n_{B_{1}}$, where $X\equiv M,B_{2}$.

\item  Coherent particles emitted via the DCPE must be coplanar with the
incoming and outgoing projectiles: strong $(\theta _{1k},\omega )$ and $
(\theta _{1k},E_{p})$ correlations, where $E_{p}$ is the total projectile
energy. 
\smallskip 

\item  The intensities as well as the absorption effects can be calculated
as in the case of generalized Cherenkov-like effects \cite{3} by using
Feynman diagrams in medium.

\item  Any two-body decay process B$_{1}$ $\rightarrow $ MB$_{2}$ in medium
via DCPE posses two limiting modes: the nuclear MCR (NMCR) mode in which the
particle M is emitted with the coherence condition $v_{Mph}\leq v_{B_{1}}$,
and the nuclear BCR (NBCR) mode with $v_{B_{2}ph}\leq v_{B_{1}}$.
\end{itemize}

\bigskip

\noindent \textsc{The electromagnetic DCPE.} To be more specific let us
consider a charged particle (e.g. $e^{\pm },\mu ^{\pm }$,p, etc.) moving in
a (dielectric, nuclear or hadronic) medium and to explore  the $\gamma $ 
coherent emission via DCPE in that media.  In the case of dielectric medium
the CR is well established phenomenon  widely used in physics and
technology.  Also, experimentally, the high energy $\gamma$-emission via
coherent bremsstrahlung (CB) is  well known as a channeling effect.  So, the
high energy component of the DCPE phenomenon can be identified  with the
CB-radiation.  Hence, more experimental and theoretical investigations are
needed since  CR and CB radiations can be described in a unified way via the
DCPE as two-component generalized  Cherenkov-like effects.  Then, the same
final interaction Hamiltonian $H_{\mathrm{fi}} $ just as in the quantum
theory of CR with some modifications of the source  fields in medium can
also describe the coherent $\gamma$-emission in the CB  sector.  Therefore,
the subthreshold rings observed experimentally can also be interpreted as
DCPE signatures since in the CR sector (Fig.1(a)) the number of photons 
emitted per time unit in the interval $(\omega ,\omega +d\omega) $ in an
nonabsorbent medium will be given by 
\begin{equation}
\frac{dN_{\gamma }}{d\omega }(DCPE)\simeq \alpha
Z_{B_{1}}^{2}v_{B_{1}}\left( 1-v^{2}_{\gamma ph}v^{2}_{B_{1}ph}\right)
\end{equation}
where $\alpha =1/137 $ is the fine structure constant and $Z_{B_{1}} $ is
the electric charge of the B$_1 $  particle.

Now, it is easy to see that the DCPE-coherence condition (\ref{Eq1d})
includes  in a general and exact form the subthreshold Cherenkov-like
radiation \cite{1} since: $v^{thr}_{B_{1}}(DCPE)=v^{thr}_{B_{1}}(CR)/\mathrm{
Re}n_{B_{1}} $. So, if $\mathrm{Re}n_{B_{1}}\geq 1,\: \mathrm{then}\:
v^{thr}_{B_{1}}(DCPE)\leq v^{thr}_{B_{1}}(CR) $.

The second major result is connected with the fact that the physical domain 
of the DCPE-radiation includes a high energy component in which the charged 
particle source is \textit{``coherently''}  stopped in medium like in the
channeling radiation process.  Hence, in the CR-case, these secondary
charged particles produce \textit{secondary rings}  in the same refractive
medium.  This DCPE-effect completely solves the mystery of the \textit{
anomalous Cherenkov radiation} \cite{1a}. In fact, we obtain a
Cherenkov-Channeling (or coherent  bremsstrahlung) radiation duality.

It is worth to note that a two-component $\gamma$-CR can explain $\gamma$
-rays emission from cosmic sources \cite{2a}. \bigskip

\noindent \textsc{The strong DCPE effects}. Now, let us consider a baryon
(e.g. p, n, $\Lambda$, $\Sigma$, etc.) moving in a (nuclear or hadronic)
medium and to explore the coherent meson emission (e.g. $\pi$, K, $\eta$,
etc.) via the strong DCPE radiation in that medium.

In the nuclear medium, two kinds of the NMCR have been intensively
investigated in details in \cite{3}, namely the \emph{nuclear pionic
Cherenkov-like radiation (NPICR)} and the nuclear kaonic Cherenkov-like
radiation (NKCR). The characteristic features of the NPICR-pions predicted
in \cite{3} are the following. (i) The NPICR-coherence condition, $
v_{ph}(\omega )\leq v $,  was found to be \textit{fulfilled in the three
energy bands}: (CB1) 190 MeV$\leq \omega \leq $ 315 MeV, for all $\pi ^{\pm
,0} $; (CB2) 910 MeV $\leq \omega \leq $ 960 MeV, only for $\pi $$^{+}$;
(CB3) $\omega \geq $ 80 GeV, for all $\pi ^{\pm ,0} $. (ii) The NPICR-pions
must be complanar with the incoming and outgoing projectile possessing
strong correlations, ($\theta_{1k},w$) and ($\theta_{1k},T_p$), where $T_p$
is the threshold kinetic energy. (iii) The \emph{NPICR differential cross
sections (DCS)} are peaked at the energy $\omega _{\mathrm{m}}=244 $ MeV for
CB1-emission band when absorption in medium is taken into account. The
CB1-peak width in the DCS is predicted to be $\Gamma _{\mathrm{m}} \leq $ 25
MeV. (iv) The NPICR-peak position is predicted to behave with energy as $
T_p^{-2}$, while the $A$-target dependence of the DCS is also predicted \cite
{3}.

Therefore, the physical domain of the \textit{low-energy component} of the
DCPE phenomenon can be identified with the NPICR-predicted pionic bands,
CB1, CB2, and CB3.  Hence, more experimental and theoretical investigations
are needed since  NPICR and NBCR radiations can be described in an unified
way via DCPE as the generalized  two-component Cherenkov-like effects with
the same interaction Hamiltonian \cite{3}.

\begin{figure}[t]
\center
{\includegraphics[bb=20 149 521 644,width=5.5cm]{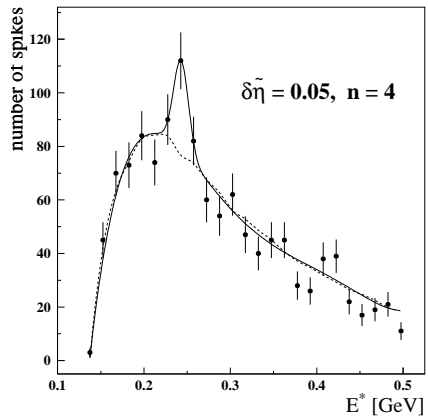}} {}
\caption{ The c.m.s. energy spectrum of dense groups (spikes) of $\pi ^{-}$s
with required multiplicity $n=4$ within a rapidity window $\delta {\stackrel{
\sim}{\eta}}=0.05$ obtained in central Mg-Mg collisions at 4.3$A$ GeV/$c$ 
\protect\cite{5,6}. The dashed and solid lines show the inclusive background
and the fitted curve, respectively. }
\label{fig:3}
\end{figure}

On the other hand, we must underline that the existence of the
NPICR-CB1-emission band has been experimentally confirmed in the studies of
dense groups, or spikes, of negative pions in central Mg-Mg collisions at 4.3
$A$ GeV/$c$ from the 2m Streamer Chamber SKM-200 (JINR, Dubna) \cite{5,6}.
In about 14420 events, the spikes have been extracted in each event from the
ordered pseudorapidities scanned with the rapidity window of the fixed size $
\delta {\stackrel{\sim}{\eta}}$ with the definite number $n$ of pions
required to hit in the window. An example of the  c.m. energy distributions
is shown in Fig.~\ref{fig:3}. So, from Fig. \ref{fig:3} and the $E^*$
-spectra at other $\delta {\stackrel{\sim}{\eta}}$ and $n$ \cite{5,6}, one
finds a significant peak in the energy spectra of emitted pions over the
inclusive background. The experimental values of the peak position, $E^*_{
\mathrm{m}}=238\pm 3(\mathrm{stat})\pm 8(\mathrm{syst})\: \mathrm{MeV}$, and
its width, $\Gamma _{\mathrm{m}}=10\pm 3(\mathrm{stat})\pm 5(\mathrm{syst})
\: \mathrm{MeV}$,  are found to be in a good agreement with the absolute
NPICR predictions for $\omega_{\mathrm{m}}$ and $\Gamma_{\mathrm{m}}$ \cite
{2,3}. It is important to note that the value of $E^*_{\mathrm{m}} $ is
similar to the position of the peak found in \cite{15} in the study of $\pi
^{+} $-production in coincidence measurements of (p,n) reactions at 0$
^{\circ} $ on C, the effect connected with NPICR \cite{3}.

\noindent Finally, we can conclude that DCPE intensities are large enough in
order to be experimentally measured in exclusive experiments by rapid
coincidence techniques. \bigskip

{}

\begin{figure}[tbp]
\centering
\hspace*{-.9cm} \resizebox*{5.7cm}{5.5cm}{\includegraphics{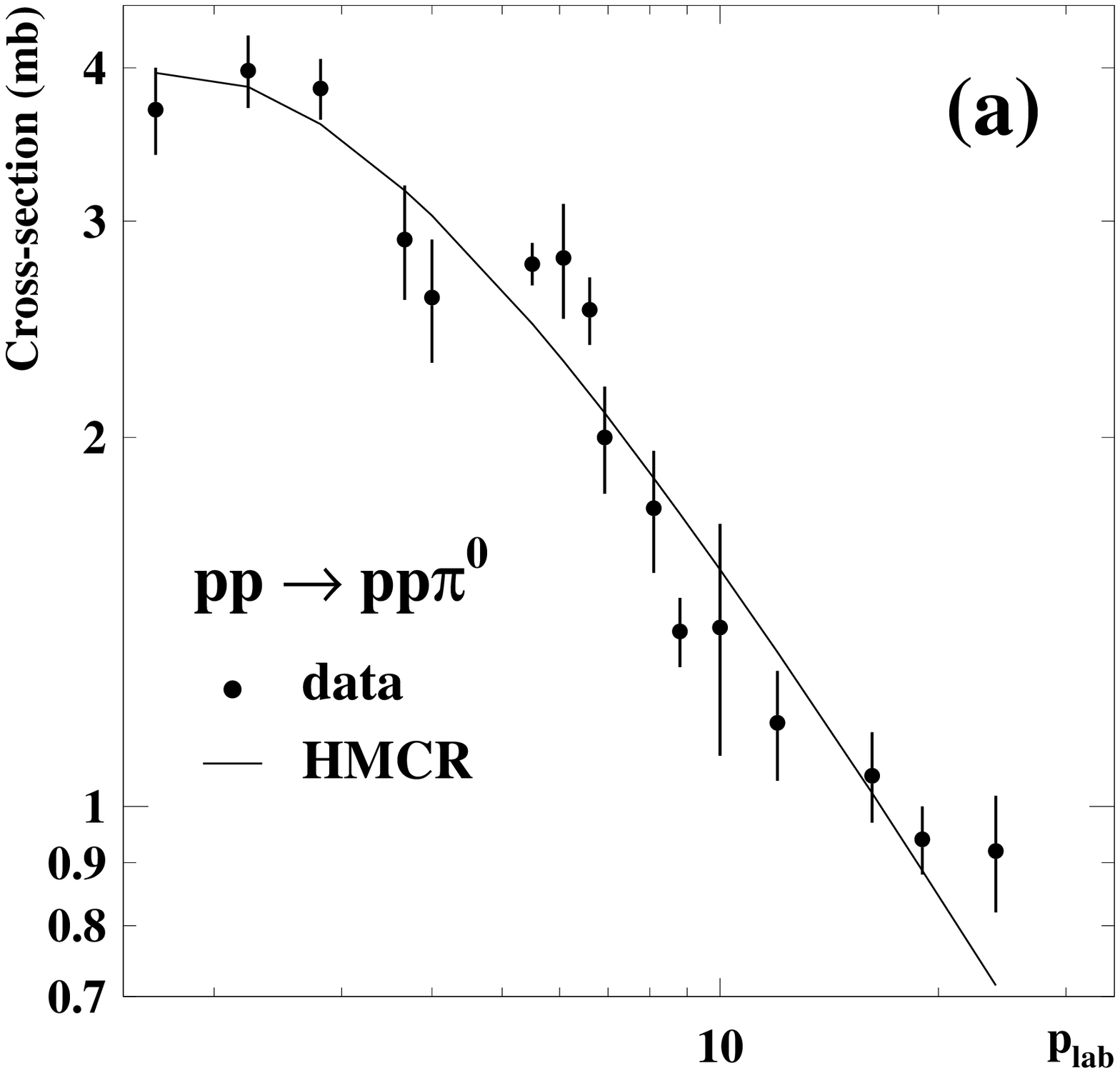} }
\hspace*{1.cm} \resizebox*{5.6cm}{5.9cm}{\includegraphics{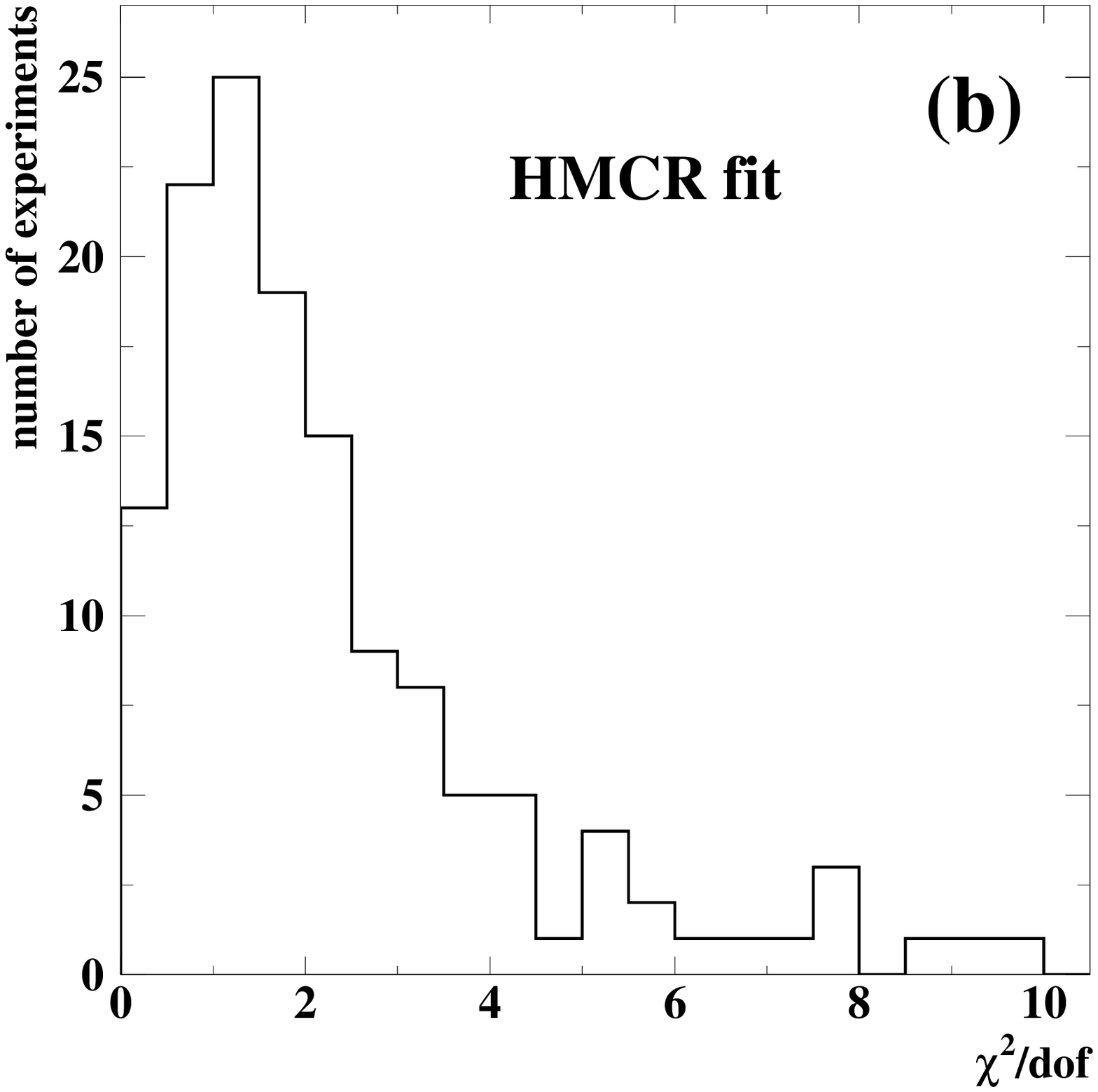} }
\vspace*{-.3cm}
\caption{ (a) Comparison between the experimental integrated cross sections
of the reaction pp$\to$pp$\pi^0$ and those predicted by the pion emission
via HPICR effect, Eq. (\ref{sig13}). (b) Experimental statistical test of
the \emph{HMCR}-\emph{mechanism dominance} in hadron-hadron collisions: the
number of reactions fitted with the HMCR for integrated cross sections vs. $
\chi ^{2}/dof $ \protect\cite{9,11}.}
\label{fig:5}
\end{figure}

\noindent \emph{\textsc{The DCPE in hadronic media.}} The mesonic
Cerenkov-like radiation in hadronic media was considered by many authors 
\cite{9,16,11}. A systematic investigation of the classical and quantum
theory of this kind of effects in hadronic media can be found in Ref. \cite
{9}. The classical variant \cite{2,9} of the Cherenkov mechanism was applied
to the study of single meson production in hadron-hadron interactions at
high energies. This variant is based on the usual assumption that hadrons
are composed from a central core in which most of the hadron mass is
concentrated surrounded by a large and more diffuse mesonic cloud (hadronic
medium). Then, it was shown [9,11] that a \emph{hadronic mesonic
Cherenkov-like radiation (HMCR)} with an mesonic refractive index given by a
pole approximation, can be able to explain the integrated cross sections of
a single meson production in hadron-hadron interactions.  Indeed, in the
pole approximation the integrated cross section reads: 
\[
\sigma _{HMCR}(v)=\sigma _{0}\, \frac{G^{2}}{4\pi }\, \, \frac{a}{Y_v} \:
\times 
\]
\begin{equation}
\left\{ 1- {\frac{1}{{Y_v}}} \left[ \frac{m}{a}+\frac{a^{2}+m^{2}}{a^{2}}
\arctan \left( \frac{aY_v-m}{mY_v+a}\right) \right] \right\}  \label{sig13}
\end{equation}

\noindent for the (pseudo)scalar mesons, and 
\begin{equation}
\sigma _{HMCR}(v)=\sigma _{0}\, \frac{G^{2}}{4\pi } \cdot \left[ \frac{
a^{2}+m^{2}}{a^{2}v^2} \arctan \left( \frac{aY_v-m}{mY_v+a}\right) - 
\frac{aY_v-m}{{Y_v}^2} \right]  
\label{sig14}
\end{equation}

\noindent for the vector mesons \cite{9,11}. Here, $v$ is the projectile
velocity, $Y_v=v/\sqrt{1-v^{2}}$, $m$ is the rest mass of the produced
meson, $G$ is the usual coupling constant, $\sigma_0=0.389$ mb, and $a
=0.350 $ GeV. To illustrate, in Fig. \ref{fig:5}(a) we present the measured
integrated cross sections of the process $\mathrm{pp\rightarrow pp\pi^{0} }$, 
compared with the HPICR-predictions. This result was very encouraging for
the extension of the HMCR analysis to other single meson production in
hadron-hadron collisions at high energies. Collecting the $\chi ^{2}/dof $s
for all 139 reactions fitted with the HMCR approach \cite{11}, we get the
surprisingly good description as shown in Fig. \ref{fig:5}(b). We must
underline that only reactions with single meson production were fitted (a
single parameter fit) with the HMCR predictions on the integrated cross
sections.

{} \vspace{0.3cm}

\textsc{Conclusions.} A new kind of coherent particle production mechanism,
called \emph{dual coherent particle emission (DCPE)}, is introduced. The
DCPE phenomenon as generalized two-component Cherenkov-like effects can  be
viewed as two body decays B$_{1}\rightarrow$ B$_{2}$M in medium.  They  are
expected to take place when the \emph{\ phase velocities of the emitted
particle $v_{Mph} $ and that of particle source $v_{B_1ph} $ satisfy the
dual coherence condition: $v_{B_{1}ph}v_{Mph}\leq 1. $} The secondary B$_{2} 
$-particles produced by the high-energy component (BCR) of the DCPE can
produce  secondary MCR-effects.  Hence, under certain circumstances the DCPE
phenomenon applied to the CR  in dielectrics (or crystals) can explain not
only subthreshold CR \cite{1} but  also the observed secondary rings (or
anomalous CR) \cite{1a}. For illustration, experimental evidences for DCPE
effects in various media are presented.

Finally, it is important to remark that more investigations of DCPE  effects
are necessary in connections with Cherenkov particle detectors  since they
may help to explain discrepancies between some experimental  results and
theoretical predictions in high-energy physics. \newpage

\noindent \textbf{\Large Acknowledgments}\newline

\noindent We would like to thank G. Altarelli for fruitful discussions. One
of the authors (D.B.I.) would like to thank TH Division for hospitality
during his stay at CERN.

\end{document}